\documentclass[11pt]{article}
\usepackage{graphicx,psfrag,epsf,color}
\usepackage{amsfonts}
\usepackage{amsmath}

\newcommand{\refequation}[1]{{Eq.(\ref{#1})}}
\newcommand{\refeq}[1]{{(\ref{#1})}}

\catcode`@=11
\newcount\@tempcntc
\def\@citex[#1]#2{\if@filesw\immediate\write\@auxout{\string\citation{#2}}\fi
  \@tempcnta\z@\@tempcntb\m@ne\def\@citea{}\@cite{\@for\@citeb:=#2\do
    {\@ifundefined
       {b@\@citeb}{\@citeo\@tempcntb\m@ne\@citea\def\@citea{,}{\bf ?}\@warning
       {Citation `\@citeb' on page \thepage \space undefined}}%
    {\setbox\z@\hbox{\global\@tempcntc0\csname b@\@citeb\endcsname\relax}%
     \ifnum\@tempcntc=\z@ \@citeo\@tempcntb\m@ne
       \@citea\def\@citea{,}\hbox{\csname b@\@citeb\endcsname}%
     \else
      \advance\@tempcntb\@ne
      \ifnum\@tempcntb=\@tempcntc
      \else\advance\@tempcntb\m@ne\@citeo
      \@tempcnta\@tempcntc\@tempcntb\@tempcntc\fi\fi}}\@citeo}{#1}}
\def\@citeo{\ifnum\@tempcnta>\@tempcntb\else\@citea\def\@citea{,}%
  \ifnum\@tempcnta=\@tempcntb\the\@tempcnta\else
   {\advance\@tempcnta\@ne\ifnum\@tempcnta=\@tempcntb \else \def\@citea{--}\fi
    \advance\@tempcnta\m@ne\the\@tempcnta\@citea\the\@tempcntb}\fi\fi}
\catcode`@=12

\begin{document}

\begin{titlepage}

\begin{flushright}
TTK-11-33\\
August 9, 2011
\end{flushright}

\vskip1.5cm
\begin{center}
\Large\bf\boldmath 
Non-local Higgs actions:
Tree-level electroweak constraints and high-energy unitarity 
\end{center}

\vspace{1cm}
\begin{center}
{\sc M.~Beneke},
{\sc P.~Knechtges},
{\sc A.~M\"uck}\\[5mm]
  {\it Institut f{\"u}r Theoretische Physik 
        und Kosmologie, RWTH Aachen,}\\
  {\it D--52056 Aachen, Germany}\\[0.5cm]
\end{center}

\vspace{2cm}
\begin{abstract}
\noindent
We consider electroweak symmetry breaking by a certain class of 
non-local Higgs sectors. Extending previous studies employing
the Mandelstam condition, a straight Wilson line is used to 
make the Higgs action gauge invariant. We show the unitarization of 
vector-boson scattering for a wide class of non-local actions, 
but find that the Wilson-line model leads to tree-level 
corrections to electroweak precision observables, which restrict 
the parameter space of the model. We also find that Unhiggs 
models cannot address the hierarchy problem, once the 
parameters are expressed in terms of low-energy observables.
\end{abstract}
\end{titlepage}

\section{Introduction}

Understanding the mechanism for electroweak symmetry breaking is one
of the major endeavours in modern particle physics. While the  
Higgs mechanism of the standard model (SM) might be a valid 
description of the physics at the TeV scale, the
associated hierarchy or little hierarchy problem provides a major 
inspiration to particle physics model building and a plethora of SM
extensions is under theoretical investigation and will be tested
at the LHC. 

Among the more exotic ideas is the suggestion that the action that 
describes the scalar sector of the standard model is non-local. 
In the spirit of unparticle physics~\cite{Georgi:2007ek}, it has been
proposed that the Higgs boson itself is an unparticle field, called 
the ``Unhiggs''~\cite{Stancato:2008mp}. The corresponding action 
reads in momentum space
\begin{equation}
S \supset \int \frac{d^4p}{(2\pi)^4}\,\tilde \phi^\dagger(p) 
\tilde F(p^2)\tilde \phi(p),
\label{eq:snonlocal1}
\end{equation}
where $\phi$ is the standard Higgs doublet field, $\tilde \phi$ 
its Fourier transform, and $\tilde F(p^2)$ takes the form 
\begin{equation}
\label{eq:unhigss_kernel}
\tilde F(p^2) = -(\mu^2-p^2-i\epsilon)^{2-d} \, 
\end{equation}
with $1\leq d < 2$. 
The Higgs field might be described by this action if it originates from 
a conformal sector and acquires a non-canonical scaling dimension $d$, 
perhaps as a consequence of some higher-dimensional 
dynamics~\cite{Falkowski:2008yr} in the context of
a soft-wall version~\cite{Cacciapaglia:2008ns,Falkowski:2008fz} of 
Randall-Sundrum models~\cite{Randall:1999ee}. The scale $\mu$ implements the
requirement that the conformal symmetry must be broken at 
low energies to avoid new light particles with electroweak 
interactions that would otherwise follow from the spectral function 
derived from $\tilde F(p^2)$. 
In this paper we will not be concerned with explaining 
the origin of \refequation{eq:snonlocal1}. Rather, we assume that 
it provides an effective description below a cut-off scale $\Lambda\gg \mu$, 
and ask whether a non-local 
Higgs interaction  of the above form provides a theoretically consistent and 
phenomenologically viable realization of electroweak symmetry 
breaking.

This question was investigated by Stancato and Terning \cite{Stancato:2008mp} 
who implemented the electroweak gauge interactions of the non-local Higgs 
boson through the Mandelstam condition~\cite{Mandelstam:1962mi}. 
To this end write \refequation{eq:snonlocal1} in position space,
\begin{equation}
\label{eq:wirkungortsraum}
S \supset \int d^4 x\, d^4 y\,  
\phi^\dagger(x) \,  F(x-y) \, W(x,y) \, \phi(y) \, ,
\end{equation}
where $F(x-y)$ is the Fourier transform of $\tilde F(p^2)$, 
and we added the Wilson line
\begin{equation}
\label{eq:Wilson_line}
W(x,y) = \mathbb{P}\exp\left( 
    -\frac{ig}{2}\int_y^x dz_\mu \,\sigma^a A^\mu_a (z) 
    -\frac{ig'}{2} \int_y^x dz_\mu B^\mu (z)\right)\,,
\end{equation}
i.e.\ the path-ordered exponential of 
$\mathrm{SU(2)}\times\mathrm{U(1)}$ gauge fields 
that connects the points $x$ and $y$ and ensures the 
gauge invariance of the action. The Mandelstam condition 
requires $[D_\mu^y W(x,y)] =0$, where $D_\mu^y$
is the usual covariant derivative. Formally Taylor-expanding 
$\tilde F(p^2)$ we then obtain from \refequation{eq:wirkungortsraum} 
\begin{equation}
S \supset  \sum_{n=0}^\infty \frac{\tilde F^{(n)}(0)}{n!} 
\int d^4 x\, \phi^\dagger(x) \, (-D^\mu D_\mu)^n \, \phi(x) \, .
\label{eq:Sseries}
\end{equation}
For $d=1$ this reduces to the standard, local, gauge-invariant 
Higgs action. The Mandelstam condition is equivalent to a generalized 
minimal-coupling prescription 
$\partial_\mu \to D_\mu$~\cite{Galloway:2008jn,Ilderton:2008ab}.
Using the Mandelstam condition to generate the gauge interactions 
of the Higgs field Stancato and Terning demonstrated that 
one obtains the correct gauge-boson masses as well as the unitarization 
of WW scattering at high energies. They also argued that the 
fine tuning of the Higgs mass is reduced, if $d$ is larger than the standard 
model (SM) value $d=1$.

However, as pointed out in \cite{Ilderton:2008ab,LewisLicht:2008mq}, 
the Mandelstam condition in its literal interpretation is not entirely 
satisfactory, since $[D_\mu^y W(x,y)] =0$ is 
solved only by field configurations that are pure gauges and therefore 
unphysical. 
More precisely, while the Mandelstam condition provides a prescription 
for evaluating $\partial_\mu^y W(x,y)$ independent of the path from 
$x$ to $y$, for any given path the Mandelstam derivative coincides 
with the usual notion of derivative only for pure gauges. 
A derivation of the series \refeq{eq:Sseries} and the
resulting generalized minimal-coupling prescription is more subtle
and may be given in terms of an operator representation of the 
action~\cite{Licht:2008km}. However, the resulting action is by no 
means uniquely fixed by gauge invariance~\cite{Ilderton:2008ab}, and 
only knowledge of the high-energy theory pins down a specific 
low-energy model (for instance, the 5D dynamics in Ref.~\cite{Falkowski:2008yr}
leads to the minimal-coupling prescription). This might be 
expected, since without the constraint of locality the action can 
depend in many ways on $D_\mu D_\nu/\mu^2$ without the need to 
introduce a scale that would be required to compensate the dimension 
of local operators. This raises the question whether the conclusions 
of Ref.~\cite{Stancato:2008mp} depend on the particular choice of the 
Mandelstam condition. In this paper, we therefore work with 
the gauge-invariant Wilson-line action \refeq{eq:wirkungortsraum} 
directly. Lorentz invariance is guaranteed by the 
straight-line path from $y$ to $x$~\cite{Ilderton:2008ab,LewisLicht:2008mv} 
in the path-ordered exponential. 

That this defines a theory different from the one based on the 
Mandelstam prescription can be exemplified by choosing $d=0$ and $\mu=0$. 
(The fact that this would lead to a quartic kinetic term need not 
concern us for the purpose of making the point.) In this case, the 
theory is local, but since the mass dimension of the Higgs field is zero 
there are several gauge-invariant operators built from covariant derivatives 
with mass dimension four which can be inserted between two Higgs fields.
Thus, the action can be of the form
\begin{equation}
\label{eq:ActionMandelstam}
\mathcal{S}\supset\int d^4x \,\,
\phi^\dagger(x) \left( c_1 (D^\mu D_\mu)(D^\nu D_\nu) + 
i g c_2 D_\mu F^{\mu\nu} D_\nu +
(i g)^2 c_3 F^{\mu\nu}F_{\mu\nu} \right)\!(x) \, \phi(x) \, ,
\end{equation}
where $D_\mu=\partial_\mu+ig T^a A^a_\mu$ denotes a generic covariant 
derivative, and $F^{\mu\nu}=[D^\mu,D^\nu]/(ig)$ the corresponding field 
strength. Minimal coupling corresponds to the choice
$c_1=-1$, $c_2=c_3=0$ of the arbitrary coefficients $c_i$. 
This choice is obviously not unique. Since partial derivatives
commute, we could equally well introduce covariant derivatives for 
$\partial_{\mu}\partial_{\nu}\partial^{\mu}\partial^{\nu}$
yielding $c_1=-1$, $c_2=1$, and $c_3=0$.
The action for the straight Wilson line is given by
\begin{equation}
\label{eq:p4action}
\mathcal{S}\supset - \int d^4x \int d^4y\,\,
\phi^\dagger(x) \, 
\left[ (\partial_y^\mu \partial_{y,\mu})(\partial_y^\nu \partial_{y,\nu}) 
\delta(x-y) \right]\! W(x,y) \phi(y) \, .
\end{equation}
Expanding the path-ordered exponential of the straight Wilson line, 
\begin{align}
W(x,y)=1 & -\frac{ig}{2}\int_0^1 \! ds_1 \, (x-y)_\mu \,
\sigma^a A^\mu_a(s_1) \\&
+\left(\frac{-ig}{2}\right)^2\int_0^1 \! ds_1 \int_0^{s_1} \! ds_2 \, 
(x-y)_\mu (x-y)_\nu \,
\sigma^a A^\mu_a(s_1)\sigma^b A^\nu_b(s_2) + \ldots \nonumber \, ,
\end{align}
it is obvious that due to the delta-function the series, when inserted into 
\refequation{eq:p4action}, terminates at 
$\mathcal{O}(g^4)$. It is a straight-forward exercise to express the 
resulting action in terms 
of covariant derivatives and field-strength tensors. One finds 
$c_1=-1$, $c_2=2/3$, and $c_3=1/6$.  
Hence, the straight Wilson line indeed leads to a different action 
compared to minimal coupling. Only for $d=1$ is the usual kinetic term 
with two derivatives unambiguously defined by gauge invariance so 
that the minimal coupling and the Wilson-line theory are equivalent. 

By focusing on the Wilson-line theory in this paper, we investigate 
another model in theory space, whose kinetic term is apparently less 
specific than minimal coupling. We shall find that
\begin{itemize}
\item contrary to the minimally coupled theory, 
there exist tree-level corrections to the $S$-parameter and 
the W mass, which put strong limits on the IR cut-off scale $\mu$, 
if $d$ is not close to the SM limit $d=1$;
\item WW scattering is unitarized for a general kernel $F$, 
but involves a more complicated interplay of different Feynman diagrams
and couplings than for the Mandelstam prescription;
\item the model does not provide a solution to the hierarchy 
problem in the following sense: when the parameters of the action 
are expressed in terms of quantities at the electroweak scale 
(such as the top and Higgs mass), 
the scale $\Lambda$ up to which the model can be valid without 
engineering large cancellations between the bare Higgs 
mass and its quantum correction cannot be raised relative to the SM. 
This is also true in the minimal-coupling theory.
\end{itemize} 

The paper is organized as follows. In Section~\ref{sec:non-local_Higgs} we 
investigate in detail the straight Wilson-line approach to a 
non-local Higgs sector, in particular electroweak symmetry breaking. 
The phenomenological consequences are studied in 
Section~\ref{sec:constraints}, and the parameter space of the 
specific Unhiggs model of \refequation{eq:unhigss_kernel} is constrained. 
In Section~\ref{sec:yukawa} we comment on the little hierarchy problem. In 
Section~\ref{sec:unitarization} we show how vector-boson scattering is 
unitarized by the non-local Higgs sector using a straight Wilson line 
for a general kernel $F(x-y)$ and comment on the Goldstone-boson
equivalence theorem. We conclude in Section~\ref{sec:summary}. The general 
structure of the interactions of the non-local Higgs field with 
gauge bosons is detailed in the Appendix.

\section{Non-local Higgs action with 
straight Wilson lines}
\label{sec:non-local_Higgs}

We consider the non-local Higgs action~\refeq{eq:wirkungortsraum},
where $F(x-y)$ is a general kernel which depends only on the coordinate
difference of the two Higgs fields in order to guarantee translational
invariance. Lorentz invariance is also assumed and made explicit in 
momentum space, since the Fourier transform ${\tilde F}(p^2)$ of $F(x-y)$
is supposed to depend only on $p^2$. However, we do not assume the 
explicit form~\refeq{eq:unhigss_kernel} unless stated otherwise. We 
only assume that it is a smooth function and leads to a sensible 
spectral function. As stated above, in the Wilson line~\refeq{eq:Wilson_line} 
the gauge fields are integrated along the straight line connecting 
the points $x$ and $y$ in order not to spoil Lorentz invariance. We use 
the standard notation for the SU(2)$_\mathrm{L}$ (U(1)$_\mathrm{Y}$) 
couplings $g$ ($g'$), and the gauge 
fields $A^\mu_a$ ($B^\mu$). The SU(2)$_\mathrm{L}$ generators are
expressed in terms of the Pauli matrices $\sigma^a$ with $a=1,2,3$.
The gauge-Higgs interactions are obtained in perturbation theory 
by expanding the path-ordered exponential, which provides interaction 
vertices of two Higgs fields with an arbitrary number of gauge fields. 
The Feynman rules are given in Appendix~\ref{se:Feynrules}.

The Higgs action is supposed to break the electroweak 
symmetry in analogy to the SM. Hence, we consider a Higgs 
potential $V$ leading to a non-zero vacuum expectation value (vev) 
$\tilde{v}$ of the Higgs doublet. Following Ref.~\cite{Stancato:2008mp} 
we make the simplest ansatz and take the potential to be local, i.e.\ 
\begin{equation}
\label{eq:higgspotential}
S \supset - \int d^4 x\, V(\phi^\dagger(x) \, \phi(x)) \, ,
\end{equation}
and parameterize the Higgs field according to
\begin{equation}
\label{eq:Higgs_doublet}
\phi = \frac{1}{\sqrt{2}} \left(
 \begin{array}{c} \hat{\pi}_2 + i\hat{\pi}_1 \\ 
  \tilde{v}+\hat{\rho} -i\hat{\pi}_3
 \end{array}\right)\, ,
\end{equation}
where $\hat{\rho}$ is the physical Higgs field, $\hat{\pi}_i$ ($i=1,2,3$) 
denote the would-be Goldstone fields, and $\tilde{v}$ is the vacuum 
expectation value of the Higgs field.
The hat indicates that the fields, as well as the vev $\tilde v$, have 
mass dimension
depending on the mass dimension of the kernel $F$, which might differ 
from one. We define scalar fields and a vev of mass dimension one via 
\begin{equation}
\label{eq:Goldstone_normalization}
(\rho,\pi,v) = \sqrt{\tilde{F}^{(1)}(0)} \;(\hat \rho,\hat\pi,\tilde v)\,,
\end{equation} 
which will turn out to be the natural normalization of the fields and 
give $v\approx 246\,$GeV from the value of the Fermi constant. 
Expanding the momentum-space action to quadratic order in the fields 
we obtain 
\begin{align}
\label{eq:higgsbilinear}
S \supset \int \frac{d^4 p}{(2 \pi)^4}\,\, & \left[ 
\frac{{\tilde F}(0)  - V^{(1)}(\tilde{v}^2/2)}{\tilde{F}^{(1)}(0)} 
\, v \, \rho(0) \, \delta^{(4)}(p)
\right. \\
& \left. \hspace*{-1.7cm} + \,\frac{1}{2} \,
\frac{{\tilde F}(p^2)  - V^{(1)}(\tilde{v}^2/2)}{\tilde{F}^{(1)}(0)} 
\left( \rho^\dagger \rho + \pi_1^\dagger\pi_1 + \pi_2^\dagger\pi_2 +
\pi_3^\dagger\pi_3 \right)(p) - 
\frac{\tilde{v}^2}{2} \frac{V^{(2)}(\tilde{v}^2/2)}{\tilde{F}^{(1)}(0)}  
\rho^\dagger(p)\rho(p) 
\right] \, ,\quad 
\nonumber
\end{align}
where the Higgs potential has been Taylor expanded around the vev. Note 
that from now on we will drop the tilde on the Fourier transformed 
fields. Since the momentum independent part of ${\tilde F}(p^2)$ 
contributes to the potential, the condition that the vev should minimize 
the potential reads
\begin{equation}
\label{eq:MinimalitaetsbedingungV}
\tilde{F}(0) - V^{(1)}(\tilde{v}^2/2) = 0 \,. 
\end{equation}
The Higgs-boson propagator is given by
\begin{equation}
\label{eq:Higgsprop}
\tilde{\Delta}_{\rho} = \frac{i \tilde{F}^{(1)}(0)}{\tilde{F}(p^2)-
\tilde{F}(0)-V^{(2)}(\tilde{v}^2/2)\tilde{v}^2} \, .
\end{equation}
Note that in general the non-local nature of the Higgs boson manifests 
itself in a continuum of states described by the cut of the Higgs 
propagator generated by the kernel $\tilde F$.

The mass  $M_H$ of the physical Higgs boson is determined by the pole 
of the propagator located at 
\begin{equation}
\label{eq:Higgsmass}
\tilde{F}(M_H^2) - \tilde{F}(0) - V^{(2)}(\tilde{v}^2/2) \tilde{v}^2 = 0 \, .
\end{equation}
We assume~\cite{Stancato:2008mp} a SM-like Higgs potential 
\begin{equation}
V(\phi^\dagger(x) \, \phi(x)) = \lambda \left[
\frac{\phi^\dagger \, \phi}{\Lambda^{2d-2}} -\frac{V^2}{2} \right]^2 \, ,
\label{eq:Higgspotential}
\end{equation}
where $\Lambda$ is the cut-off of the theory, which is inserted here for
dimensional reasons because of the unusual mass dimension $d$ of the 
Higgs field, and the unparticle motivated form~\refeq{eq:unhigss_kernel} 
of the non-local kernel. In this specific case, one finds 
\begin{equation}
v^2 = {\tilde v}^2\tilde{F}^{(1)}(0)={\tilde v}^2(2-d)\mu^{2-2d}
\label{eq:vevtilde}
\end{equation}
and
\begin{equation}
\label{eq:Higgs_mass_unexpanded}
M_H^2= \mu^2 - \mu^{2} \left[ 1 - 
\frac{2 \lambda}{2-d}
\frac{v^2}{\mu^2} 
\left(\frac{\mu}{\Lambda}\right)^{4d-4} \right]^{\frac{1}{2-d}} \, .
\end{equation}
We would like $M_H^2$ to be positive and therefore assume parameters 
such that the second term in square brackets is smaller than 1. If it 
is much smaller, $M_H$ is approximately  
\begin{equation}
\label{eq:approximate_unhiggs_mass}
M_H^2 \approx \frac{2 \lambda v^2}{(2-d)^2} 
\left(\frac{\mu}{\Lambda}\right)^{4d-4} \, ,
\end{equation}
which shows that $\Lambda/\mu$ cannot be too large if the Higgs 
self-interaction is to remain perturbative, since otherwise $M_H$ would 
be too small. We also note that \refequation{eq:MinimalitaetsbedingungV} 
reads
\begin{equation}
\left(\frac{\mu}{\Lambda}\right)^{4d-4} \frac{\lambda v^2}{2-d} 
= \left(\frac{\mu}{\Lambda}\right)^{2d-2} \lambda V^2 - \mu^2 
\approx \frac{2-d}{2}\,M_H^2 \,,
\end{equation}  
which requires fine-tuning  of the unrelated parameters $\mu$ and $V$, 
if $\mu$ is much larger than the electroweak scale. 
 
As in the SM, the spontaneous symmetry breaking results in 
massive gauge-boson fields and the mixing of the gauge fields with the 
Goldstone modes. The latter can be removed by a suitable gauge-fixing 
condition. The mixing terms are obtained by expanding the Wilson line in 
\refequation{eq:wirkungortsraum} to first order in the gauge fields. Employing
translational invariance, one finds 
\begin{align}
S_{\mathrm{Mixing}} = & \,\frac{v}{4}\frac{1}{\tilde{F}^{(1)}(0)} 
\int \frac{d^4p}{(2\pi)^4} 
\left[g \pi^{a\dagger}(p) A^a_\mu (p) - g' \pi^{3\dagger}(p) B_\mu (p)
\right]\\
&  \times\int d^4x \, \, F(x) \left(1 - \exp (ipx) \, \right) 
\int_0^x dz^\mu \exp (-ip z)\, .
\nonumber
\end{align}
Parameterizing the straight Wilson line from 0 to $x$ by $z^\mu=x^\mu s$ 
and using 
\begin{equation}
\label{equ:derivativetrick}
\int d^4x \, \, F(x) \exp (-iqx) \, \int_0^1 ds \, x^\mu \exp (-is p x) = 
\left. i\, \int_0^1 ds \, \frac{\partial}{\partial q'_\mu} {\tilde F}(q'{}^2)
\right|_{{q'}^\mu=q^\mu+p^\mu s},
\end{equation}
the integrands can be written as total derivatives with respect to $s$ 
such that the Wilson-line integral can be performed. The mixing term 
then reads
\begin{equation}
S_{\mathrm{Mixing}} = \frac{v}{2}\int \frac{d^4p}{(2\pi)^4} 
\left[g \pi^{a\dagger}(p) A^a_\mu (p) - g' \pi^{3\dagger}(p) B_\mu (p)
\right] \frac{i p^\mu}{p^2} 
\frac{{\tilde F}(p^2) - {\tilde F}(0)}{\tilde{F}^{(1)}(0)} \,.
\end{equation}
To remove these terms the gauge-fixing terms
\begin{align}
\label{eq:GF}
S_{\mathrm{GF}} =& 
- \frac{1}{2\xi} \int \frac{d^4p}{(2\pi)^4}  
\left(F_a^\dagger(A_a,\pi_a)F_a(A_a,\pi_a) + 
F^\dagger(B,\pi_3)F(B,\pi_3)\right)
\frac{\tilde{F}(p^2)-\tilde{F}(0)}{p^2 \tilde{F}^{(1)}(0)} \,
\end{align}
are added to the action, where
\begin{align}
\label{eq:gauge_fixing_condition}
F_a(A_a,\pi_a) & = p_\mu A^\mu_a(p) - \xi \frac{ivg}{2}  \pi_a(p) \quad 
\mathrm{and} \quad
F(B,\pi_3) = p_\mu B^\mu(p) +\xi \frac{ivg'}{2} \pi_3 (p)\,.
\end{align}
The mixing terms are equivalent to the ones obtained in 
Ref.~\cite{Stancato:2008mp} with the Mandelstam condition rather than 
the straight Wilson line. This is because the mixing terms stem  
from the coupling of a single gauge boson to the Higgs fields 
which is completely determined by a Ward identity of the underlying 
gauge symmetry. The above gauge fixing implies  
the Goldstone-boson propagators
\begin{equation}
\Delta_{\pi_{1,2}} = \frac{i}{p^2-\xi \hat{m}^2_W} \, 
\frac{p^2\,\tilde{F}^{(1)}(0)}{\tilde{F}(p^2)-\tilde{F}(0)} 
\quad \mathrm{and} \quad
\Delta_{\pi_3} = \frac{i}{p^2-\xi \hat{m}^2_Z } \, 
\frac{p^2\,\tilde{F}^{(1)}(0)}{\tilde{F}(p^2)-\tilde{F}(0)} \, ,
\end{equation}
where $\hat{m}^2_W = g^2 v^2/4$ and $\hat{m}^2_Z = (g^2+g'^2)\,v^2/4$. 
The Goldstone-boson propagators are also the same as with
the minimal-coupling prescription.

In contrast to the mixing terms, the coupling of two gauge bosons to 
the Higgs fields depends on the way gauge invariance is realized. Hence, 
the investigation how the gauge bosons acquire mass shows that the 
Mandelstam prescription \refeq{eq:Sseries} and the action based on 
the Wilson line \refeq{eq:wirkungortsraum} are different models with 
different phenomenological consequences.
The terms resulting in non-zero gauge-boson masses arise from expanding 
\refequation{eq:wirkungortsraum} to second order in the gauge fields when the 
Higgs field assumes its vev. In terms of the physical mass eigenstates,
i.e.\ the W- and Z-boson fields, they read
\begin{align}
S_{\mathrm{Mass}} =& - \int \frac{d^4p}{(2\pi)^4} 
\left[ \frac{\tilde{v}^2g^2}{4} W^+_\mu(p) W^-_\nu(p)+
\frac{\tilde{v}^2}{8} (g^2+g'^2) Z^{\dagger}_\mu(p) Z_{\nu}(p)\right] 
\\&\times		
\left[\,\int d^4x\, F(x) \int_0^x dz^\mu \int_0^z dz'^\nu \exp(ip(z-z'))
\right] \, ,
\nonumber
\end{align}
where the integration boundaries of the line integrals reflect the 
path ordering of the Wilson line. Parameterizing the integrals 
as before and using relations analogous to \refequation{equ:derivativetrick}, 
the additional bilinear terms in the gauge fields that arise from 
spontaneous symmetry breaking are 
\begin{align}
S_{\mathrm{Mass}} = - & 
\int \frac{d^4p}{(2\pi)^4} 
\left[\frac{\tilde{v}^2g^2}{4} W^+_\mu(p) W^-_\nu(p)+
\frac{\tilde{v}^2}{8} (g^2+g'^2) Z^{\dagger }_\mu(p) Z_{\nu}(p)\right] 
\\& \times		
\left[\,\left(g^{\mu\nu}-\frac{p^\mu p^\nu}{p^2}\right) 
\left[\frac{\tilde{F}(p^2)-\tilde{F}(0)}{p^2} -
2 \int_0^1 ds \tilde{F}^{(1)}(p^2s^2)\right] 
- \frac{p^\mu p^\nu}{p^2} \, \frac{\tilde{F}(p^2)-\tilde{F}(0)}{p^2}\,\right] 
\, .
\nonumber
\end{align}
This result can also be derived from the general interaction 
terms in Appendix~\ref{se:Feynrules}.
In the SM, where $\tilde{F}(p^2)=p^2$ and 
$\tilde v^2=v^2 [\tilde{F}^{(1)}(0)]^{-1} =v^2$, 
one recovers the usual mass terms for the gauge fields. For the 
non-local Higgs, there is a non-trivial momentum dependence which is 
similar to a self-energy that normally arises only through loops. 
In the limit $p^2 \to 0$, one recovers the SM momentum-space action, 
hence in low-energy experiments (far below the electroweak scale), 
the non-local Higgs mimics the SM electroweak symmetry breaking precisely. 
However, the full gauge-boson propagators differ from the SM and read
\begin{align}
\label{eq:gauga_boson_propagators}
\Delta_\mathrm{V}^{\mu\nu} (p) =& \,\frac{i}{p^2 -
\tilde{m}_\mathrm{V}^2 \left(2 \int_0^1 ds 
\tilde{F}^{(1)}(p^2s^2)-\frac{\tilde{F}(p^2)-\tilde{F}(0)}{p^2}\right)}
\left(-g^{\mu\nu}+\frac{p^\mu p^\nu}{p^2}\right)
\\& - \frac{i}{p^2 - \xi \hat{m}_\mathrm{V}^2} \, 
\frac{\xi p^2 \tilde{F}^{(1)}(0)}{\tilde{F}(p^2)-\tilde{F}(0)}
\, \frac{p^\mu p^\nu}{p^2}\, ,
\nonumber 
\end{align}
where $\tilde{m}_\mathrm{V}^2=g^2\tilde{v}^2/4$ 
($(g^2+{g'}^2)\,\tilde{v}^2/4$) in the W-boson (Z-boson) propagator. 
The longitudinal parts of the propagators are equal to 
the corresponding Goldstone-boson propagators up to a factor of $\xi$, 
exactly as in the SM. However, the physical gauge-boson masses 
$m_\mathrm{V}$, i.e.\ the pole of the transverse parts of the propagators, 
are shifted with respect to the 
corresponding SM relation and can be determined by solving the equation
\begin{equation}
m_\mathrm{V}^2 = \tilde{m}_\mathrm{V}^2 \left[\,2 \int_0^1 ds\,
\tilde{F}^{(1)}(m_\mathrm{V}^2s^2)-
\frac{\tilde{F}(m_\mathrm{V}^2)-\tilde{F}(0)}{m_\mathrm{V}^2}\right].
\label{eq:WilsonLineMasses} 
\end{equation}
The shift of the vector-boson masses affects electroweak precision 
observables as discussed in the subsequent section. We note that a 
minimally coupled non-local Higgs does not modify the transverse 
gauge-boson propagator~\cite{Stancato:2008mp}, and therefore leaves the 
gauge-boson masses unchanged. This is a crucial difference between 
the straight Wilson line and general non-local theories as compared 
to the ones obtained from the Mandelstam prescription.   

\section{Electroweak constraints}
\label{sec:constraints}

To constrain the non-local Higgs sector as introduced in the previous 
section by electroweak precision tests, it is convenient to represent 
the additional momentum-dependent terms as self-energy corrections 
to the gauge-boson propagators and to write the transverse propagator 
in the form
\begin{equation} 
\label{eq:propagator_with_self_energy}
\Delta_\mathrm{V}^{T\,\mu\nu} (p) = 
\frac{i}{p^2 - \hat{m}_\mathrm{V}^2 
- \Pi^\mathrm{new}_\mathrm{V}(p^2)} 
\left(-g^{\mu\nu}+\frac{p^\mu p^\nu}{p^2}\right), 
\end{equation}
with
\begin{equation}
\label{eq:definition_self_energy}
\Pi_\mathrm{V}^{\mathrm{new}}(p^2) =  
- \tilde{m}_\mathrm{V}^2 \left[\tilde{F}^{(1)}(0) + 
\frac{\tilde{F}(p^2)-\tilde{F}(0)}{p^2}  -
2 \int_0^1 ds\, \tilde{F}^{(1)}(p^2s^2)\right] \, .
\end{equation}
Note $\Pi^\mathrm{new}_\mathrm{V}(0)=0$. Thus, the ``new physics'' 
contribution vanishes at low energies as discussed above. The leading 
term in the small-momentum 
expansion (valid e.g.\ for $p^2\ll \mu^2$ for the specific 
choice~\refeq{eq:unhigss_kernel} of the kernel) is given by 
\begin{equation}
\Pi_\mathrm{V}^{\mathrm{new}}(p^2) =  \frac{1}{6}\,
p^2 \tilde{m}_\mathrm{V}^2 \,\tilde{F}^{(2)}(0)= 
\frac{1}{6} \, p^2 \hat{m}_\mathrm{V}^2\,
\frac{\tilde{F}^{(2)}(0)}{\tilde{F}^{(1)}(0)} \approx \, 
\frac{1}{6} \, p^2 m_\mathrm{V}^2\,
\frac{\tilde{F}^{(2)}(0)}{\tilde{F}^{(1)}(0)} \, .
\end{equation}
Using the conventions of the particle data group for the Peskin-Takeuchi 
parameters \cite{Peskin:1991sw} $S$, $T$, and $U$, together with 
the fact that there are no new-physics contributions in the photon 
self-energy and photon--Z-mixing in the non-local Higgs model, we find
\begin{align}
\alpha T & = \frac{\Pi^\mathrm{new}_W(0)}{m_W^2} - 
\frac{\Pi^\mathrm{new}_Z(0)}{m_Z^2} = 0 \, , \\
\frac{\alpha}{4 s_\mathrm{W}^2 c_\mathrm{W}^2} S & = 
\frac{\Pi^\mathrm{new}_Z(m_Z^2)-\Pi^\mathrm{new}_Z(0)}{m_Z^2} = 
\frac{\Pi^\mathrm{new}_Z(m_Z^2)}{m_Z^2} \approx 
\frac{1}{6} m_Z^2 \,\frac{\tilde{F}^{(2)}(0)}{\tilde{F}^{(1)}(0)}
\,, \\
\frac{\alpha}{4 s_\mathrm{W}^2 } (S+U) & = 
\frac{\Pi^\mathrm{new}_W(m_W^2)-\Pi^\mathrm{new}_W(0)}{m_W^2} = 
\frac{\Pi^\mathrm{new}_W(m_W^2)}{m_W^2} \approx
\frac{1}{6} m_W^2 \,\frac{\tilde{F}^{(2)}(0)}{\tilde{F}^{(1)}(0)} 
\approx \frac{\alpha}{4 s_\mathrm{W}^2 } S \, ,
\end{align}
where $\alpha=\alpha(m_Z)$ is the fine-structure constant and 
$\mathrm{s}_\mathrm{W}$ ($\mathrm{c}_\mathrm{W}$) the sine (cosine) 
of the weak mixing angle. Thus $T=0$ and $U\ll S$ in the present 
model. For the specific non-local kernel~\refeq{eq:unhigss_kernel}
the result is given by
\begin{align}
\frac{\alpha }{4 s_\mathrm{W}^2 c_\mathrm{W}^2} \,S = 
\frac{d-1}{6} \, \frac{m_Z^2}{\mu^2} 
+ {\cal O}\left(\frac{m_Z^4}{\mu^4}\right)
\qquad\quad 
\frac{\alpha}{4 s_\mathrm{W}^2} \,U = 
{\cal O}\left(\frac{m_Z^2(m_Z^2-m_W^2)}{\mu^4}\right)
\,,
\label{eq:S}
\end{align}
Thus the model yields a tree-level correction to the $S$-parameter that 
goes to zero for large $\mu$, as well as in the SM limit $d\to 1$.

\begin{figure}
\begin{center}
\includegraphics[width=8cm]{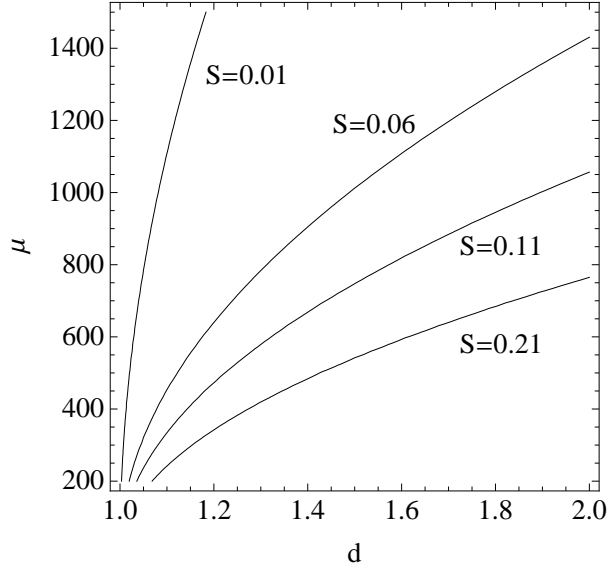}
\end{center}
\vspace*{-0.5cm}
\caption{\label{fi:S_parameter} Contours for different values of $S$ 
in the plane of the model parameters $d$ and $\mu$.}
\end{figure}

A global fit to electroweak precision measurements (assuming $U=0$) 
yields $S=0.03 \pm 0.09$ for a 
light Higgs boson \cite{Nakamura:2010zzi}. Due to a 
strong correlation with $T$ (87\%), the constraint on $S$ for a model 
with $T=0$ is even stronger, $S=-0.04 \pm 0.05$. 
In Figure~\ref{fi:S_parameter}, we show contours of $S$ in the plane of  
the two parameters $d$ and $\mu$ of the model using the approximate 
expression (\ref{eq:S}). The contours correspond to the $1\sigma$, 
$2\sigma$,  $3\sigma$, and $5\sigma$ upper limits on $S$, such that 
the region below the curve is excluded with the given significance. 
We conclude that the existence of tree-level correction forces $\mu$ 
to the TeV scale, if $d$ is not close to the SM value $d=1$.

An alternative (but not independent) constraint is obtained 
by computing the shift of the W mass, expressed in terms of the 
best measured electroweak input parameters, the fine-structure constant 
$\alpha$ (at the scale $m_Z$), the Fermi constant $G_\mathrm{F}$ and the
Z-boson mass. In terms of the 
self-energy~\refeq{eq:definition_self_energy}, the pole masses of the 
gauge bosons are given by
\begin{equation}
m_\mathrm{V}^2 = \hat{m}_\mathrm{V}^2 
\left( 1 + \frac{\Pi^{\mathrm{new}}_{V} 
(m_\mathrm{V}^2)}{\hat{m}_\mathrm{V}^2} \right)\,.
\end{equation}
To obtain the W mass we must express $g$ and $v$ in $\hat m_W^2=g^2 v^2/4$ 
in terms of $\alpha$, $m_Z$ and $G_\mathrm{F}$. Since the new 
contribution is a tree-level effect, we may use SM tree-level relations 
to compute the mass {\em shift}. We first define 
$\hat{c}_\mathrm{W}^2$ by 
\begin{equation}
\hat{m}_W^2= \frac{g^2}{g^2+g'{}^2} \,\hat{m}_Z^2 \equiv  
\hat{c}_\mathrm{W}^2 \hat{m}_Z^2\,. 
\end{equation}
To first order in the self energies and exploiting $U\approx 0$, we 
obtain 
\begin{equation}
m_W^2= \hat{c}_\mathrm{W}^2 m_Z^2 
\left( 1 - s_\mathrm{W}^2 
\frac{\Pi^{\mathrm{new}}_{Z} (m_Z^2)}{m_Z^2}
 \right)\, .
\label{eq:dmWder1}
\end{equation}
To relate $\hat{c}_\mathrm{W}$ to the input parameters we 
define $c_\mathrm{W}$ via
\begin{equation}
G_\mathrm{F}=\frac{\pi \alpha}{\sqrt{2}} 
\frac{1}{s_\mathrm{W}^2 c_\mathrm{W}^2 m_Z^2} = 
\frac{1}{\sqrt{2} v^2}\,
\end{equation}
and find 
\begin{equation}
c_\mathrm{W}^2=\hat{c}_\mathrm{W}^2 \left( 1 + 
\frac{s_\mathrm{W}^2}{c_\mathrm{W}^2-s_\mathrm{W}^2} 
\frac{\Pi^{\mathrm{new}}_{Z} (m_Z^2)}{m_Z^2} 
\right) \, .
\end{equation}
Substituting into \refequation{eq:dmWder1} we obtain the desired relation 
\begin{equation}
m_W^2=c_\mathrm{W}^2 m_Z^2 \left( 1 -
\frac{2 s_\mathrm{W}^2 c_\mathrm{W}^2}{c_\mathrm{W}^2-s_\mathrm{W}^2}\,
\frac{\Pi^{\mathrm{new}}_{Z} (m_Z^2)}{m_Z^2} \right) \,. 
\end{equation}
Since $m_W^{\rm SM} = m_W - \Delta m_W = c_\mathrm{W} m_Z$, we obtain 
the mass shift\footnote{In terms of the quantity $\Delta r$, the 
following result corresponds to
\begin{displaymath}
\frac{m_Z^2-m_W^2}{m_Z^2}\,\frac{m_W^2}{m_Z^2} = \frac{\pi\alpha}
{\sqrt{2} G_F m_Z^2}\,(1+\Delta r)
\qquad \mbox{with}\qquad 
\Delta r = \frac{\alpha}{2 s_\mathrm{W}^2} \, S\,.
\end{displaymath}}
\begin{equation}
\frac{\Delta m_W}{m_W} = - 
\frac{s_\mathrm{W}^2 c_\mathrm{W}^2}{c_\mathrm{W}^2-s_\mathrm{W}^2}\,
\frac{\Pi^{\mathrm{new}}_{Z} (m_Z^2)}{m_Z^2} 
= -\frac{\alpha}{4 (c_\mathrm{W}^2-s_\mathrm{W}^2)}\,S\,,
\end{equation}
where we expressed the result in terms of the $S$-parameter predicted 
in the model. Numerically, this turns into
\begin{equation}
\Delta m_W \approx -15\,\mbox{MeV}\times\frac{S}{0.05}\,.
\end{equation}
Since $S$ is positive, see \refequation{eq:S}, the non-local Higgs model 
with a straight Wilson line predicts a negative contribution to the 
W mass. The direct and indirect W-mass measurements 
give $m_W=(80.399\pm 0.023)\,$GeV \cite{Nakamura:2010zzi}. Since the 
error on the $S$-parameter is 0.05, which corresponds to a $15\,$MeV shift, 
we conclude that the W-mass observable alone is less constraining than 
the combination of observables that goes into $S$, as should be expected, 
but only slightly so. Even for small Higgs masses
the measured W mass is about 1-2 $\sigma$ {\em larger} than the value 
predicted in the SM, which cannot be accommodated in the present model.

Note that the constraints from electroweak precision observables for 
$d$ and $\mu$ discussed above are absent in the minimally coupled 
version of the non-local Higgs sector, which does not result in tree-level 
corrections to the transverse gauge-boson propagators.

\section{Higgs self-energy and the hierarchy problem}
\label{sec:yukawa}

The Higgs self-energy potentially receives large quantum corrections 
from the Higgs self-interaction and the Yukawa coupling to the top 
quark. In this section we discuss the cut-off dependence of the leading 
quantum correction adopting the kernel 
$\tilde F(p^2) = -(\mu^2-p^2-i\epsilon)^{2-d}$ 
from \refequation{eq:unhigss_kernel} as in Ref.~\cite{Stancato:2008mp}.
The Yukawa interaction of the physical Higgs field is 
\begin{equation}
\mathcal{L}_\mathrm{Y} = 
- \frac{1}{\sqrt{2}} \frac{\lambda_t}{\Lambda^{d-1}} \,\hat{\rho} \bar{t} t \,,
\end{equation}
where the cut-off scale has been introduced so that the Yukawa coupling 
$\lambda_t$ is dimensionless.
The Yukawa coupling can be expressed in terms of the observable 
top-quark mass and vev as 
\begin{equation}
\lambda_t =  \sqrt{2-d} \,\frac{\sqrt{2} m_t}{v} 
\left( \frac{\Lambda}{\mu} \right)^{d-1} \, .
\end{equation}
Since $\sqrt{2} m_t/v\approx 1$ is fixed by known low-energy data, 
and $\Lambda \gg \mu$, the numerical value of $\lambda_t$ tends 
to be larger than in the SM. This reflects the fact that the Yukawa 
coupling is now irrelevant and hence has to be large to produce the large 
value of the top-quark mass.

At the one-loop level, adding the Higgs self-energy 
to \refequation{eq:Higgsmass}, the Higgs mass is determined from
\begin{equation}
\label{eq:Higgsmass_1-loop}
\tilde{F}(M_H^2) - \tilde{F}(0) - V^{(2)}(\tilde{v}^2/2) 
\tilde{v}^2 - \Pi_H^{\mathrm{1-loop}}(M_H^2)= 0 \, .
\end{equation}
Since the Higgs mass is expected to be of the order of $\mu$ or smaller,
to avoid excessive fine-tuning, the self-energy contribution should be 
smaller than the tree-level term or at least $\mu^{4-2d}$ from 
$\tilde{F}(0)$. The top-loop contribution to the Higgs self-energy 
is estimated by 
\begin{equation}
\left| \Pi_H^{\mathrm{top}}(M_H^2) \right| = 
\frac{3 |\lambda_t|^2}{8 \pi^2} \,\Lambda^{4-2d} 
= \frac{3}{4 \pi^2} \frac{m_t^2}{v^2} \,(2-d) \Lambda^2 \mu^{2-2d} 
\stackrel{!}{<} \mu^{4-2d} \, .
\label{eq:toploop}
\end{equation}
For fixed Yukawa coupling the power of $\Lambda$ is reduced relative 
to the quadratic cut-off sensitivity of the SM~\cite{Stancato:2008mp}. 
However, another way to look at the hierarchy problem is to ask 
up to which scale a given theory (the SM, the non-local Higgs model ...) 
can be an effective description without engineering large cancellations 
between the bare Higgs mass and its quantum correction, {\em 
keeping the low-energy parameters (i.e. the world as we know it) 
fixed.} We then see from \refequation{eq:toploop} 
that after eliminating the Yukawa coupling in 
favour of $m_t$, which is known and fixed (contrary to $\lambda_t$ itself),  
the cut-off dependence is still quadratic and nothing is gained 
with respect to the little hierarchy problem as phrased above unless $d$ 
is very close to the singular limiting value $d=2$. Similarly, for the 
one-loop self-energy from the quartic self-coupling of the physical Higgs
field, we find 
\begin{equation}
\left| \Pi_H^{\mathrm{self}}(M_H^2) \right|= 
\frac{3 \lambda}{16 \pi^2 d} \Lambda^{4-2d}
\approx 
\frac{3 (2-d)^2}{32 \pi^2 d}\left(\frac{M_H}{v}\right)^2
\Lambda^{2 d}\mu^{4-4 d}
\stackrel{!}{<} \mu^{4-2d} \, ,
\end{equation}
using the approximate expression \refequation{eq:approximate_unhiggs_mass}
in the second equation. Hence, after trading $\lambda$ for the 
physical Higgs mass, the cut-off dependence is even more severe 
than in the SM, again unless $d$ is very close to 2. 

We therefore conclude that the non-local Higgs theory suffers from the 
same ultraviolet sensitivity problems as the SM Higgs sector. 
The difference compared to Ref.~\cite{Stancato:2008mp} arises from the 
fact that in Ref.~\cite{Stancato:2008mp} the couplings 
$\lambda_t$ and $\lambda$ were not eliminated in favour of the top 
and Higgs masses. In this case the low-energy parameters vary as 
$\Lambda$ changes and the hierarchy problem is interpreted as the 
sensitivity of the low-energy parameters to these changes. 

\section{\boldmath Unitarization of WW scattering}
\label{sec:unitarization}

It is a well-known fact that the Higgs boson in the SM plays a crucial role
in the unitarization of the scattering of electroweak gauge bosons. Without 
a Higgs boson the corresponding amplitude would grow with energy and, hence, 
necessarily violate tree-level unitarity at high energies. Adding the amplitude
for the exchange of the Higgs boson cancels all terms growing with energy at 
energies above the Higgs-boson mass. Thus, tree-level unitarity allows to 
derive upper bounds on this mass which is otherwise a free parameter 
of the SM.

Introducing a non-local Higgs sector poses the question if
the unitarization of gauge-boson scattering still holds.
Since it is ultimately a property of the gauge structure of the model, 
one might expect that unitarization should not be spoiled by the 
non-locality if implemented in a gauge-invariant way. For the Mandelstam 
prescription this has been shown explicitly at tree-level
in the scattering of W bosons~\cite{Stancato:2008mp}. Here, we extend 
these considerations to the case of the straight Wilson line, which 
guarantees gauge invariance for general kernels. We demonstrate that  
unitarization follows from a very non-trivial cancellation between different 
classes of diagrams. As discussed in Section~\ref{sec:comparison} and 
explained in Section~\ref{sec:equivalence_theorem} the leading high-energy 
behaviour of the scattering amplitude is the same for both 
gauge-invariant realizations of the action. However, the intermediate 
steps and even the contributing diagrams differ considerably.

\subsection{High-energy behaviour with straight 
Wilson lines}

In this section, we show that all potentially unitarity-violating 
terms in longitudinal W-boson scattering 
\begin{equation}
W^+_\mathrm{L}(q_1)W^-_\mathrm{L}(q_2) \to 
W^+_\mathrm{L}(q'_1)W^-_\mathrm{L}(q'_2)
\end{equation}
cancel in the sum of all Feynman diagrams. In the following, all  
amplitudes will be given in terms of incoming momenta $q_i$, 
where $i=1,\ldots4$, and for notational convenience we choose 
$q'_1=-q_4$ and $q'_2=-q_3$. We will keep all terms
which grow with energy. Furthermore, we keep also terms constant in energy 
which do not vanish in the limit of vanishing gauge couplings, after 
expressing the gauge-boson masses in terms of the gauge couplings and 
the Higgs vev. Note that due to the self-energy in the transverse
gauge-boson propagators $\Pi^{\mathrm{new}}_\mathrm{V}$ the residues of the
poles are not equal to one, leading to non-trivial $Z$-factors in the 
calculation of matrix elements. However, these global factors do not 
interfere with the cancellation of potentially unitarity-violating terms 
and differ from one only at $\mathcal{O}(g^2)$. They do not affect our 
results at the precision we are aiming at.

\begin{figure}
\begin{center}
\includegraphics[scale=.5]{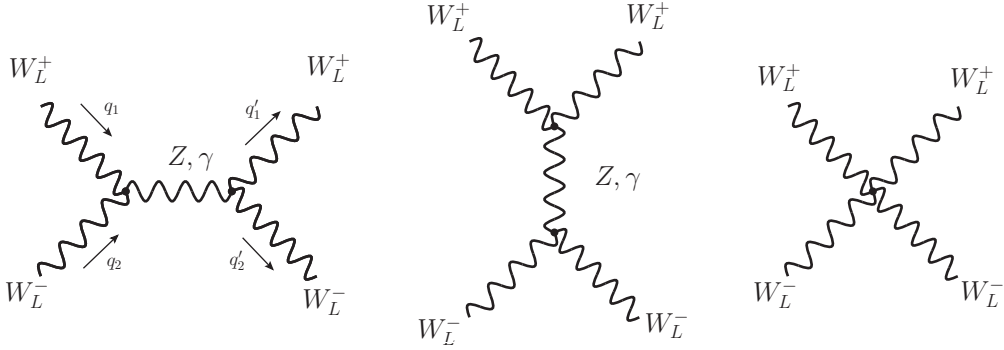}
\end{center}
\caption{\label{fig:SM_gauge_boson_graphs} Feynman graphs from the 
gauge sector which contribute to tree-level W-boson scattering.}
\end{figure}

First, we consider the diagrams shown in 
Figure~\ref{fig:SM_gauge_boson_graphs}.
Since the corresponding Feynman rules for the vertices are derived 
from the gauge sector, which is unchanged with respect to the 
SM, only the modified gauge-boson propagators 
\refeq{eq:gauga_boson_propagators} lead to a modification of the well-known 
SM result. Concerning the leading terms, growing with energy like $s^2$, 
the SM result is unchanged, i.e.\ these terms cancel in 
the sum of the three diagrams. As expected, the gauge-dependent
pieces of the propagator do not contribute and the pure gauge-sector
amplitude $\mathcal{M}_\mathrm{GS}$ is given by
\begin{equation}
\mathcal{M}_\mathrm{GS} =
\frac{ig^2}{4m_W^2} 
\left[s\left(4-2\gamma(t)-\gamma(s)\right) +
t\left(4-2\gamma(s)-\gamma(t)\right)\right] + \mathcal{O}(s^0) \, ,
\label{eq:wwgauge}
\end{equation}
where 
\begin{equation}
\gamma(s) = \frac{\hat{m}_W^2}{m_W^2} 
\left( 1 + 
\frac{\Pi^{\mathrm{new}}_Z(s)}{\hat{m}_Z^2} \right)
\end{equation}
and $\Pi^{\mathrm{new}}_\mathrm{V}(s)$ is the self-energy contribution 
defined in \refequation{eq:definition_self_energy}. 
For the Mandelstam prescription, 
the self-energy vanishes and the first factor reduces to unity, 
in which case \refequation{eq:wwgauge} 
equals the SM result~\cite{Stancato:2008mp}. However, 
for the straight Wilson line, the SM result is modified as above.

\begin{figure}
\begin{center}
\includegraphics[scale=.5]{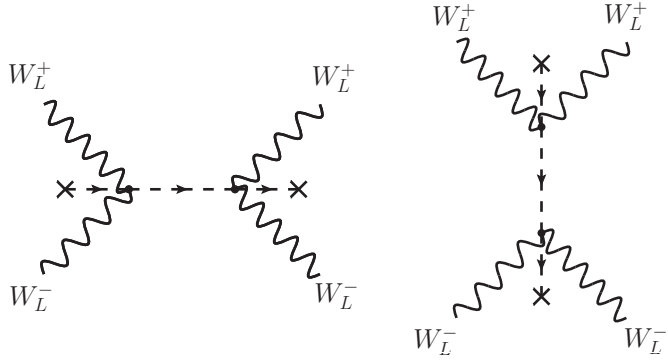}
\end{center}
\caption{\label{fig:SM_Higgs_graphs} Feynman graphs for
Higgs-boson exchange in analogy to the SM. The crosses indicate that a
Higgs boson has been replaced by its vev. }
\end{figure}

In the SM longitudinal W-boson scattering is unitarized by Higgs-boson
exchange. The corresponding Feynman graphs are shown in 
Figure~\ref{fig:SM_Higgs_graphs}. The relevant vertex function for the 
coupling of a single Higgs boson $\rho$ and two W bosons can be derived 
with the techniques described in Section~\ref{sec:non-local_Higgs} or 
from the general result in Appendix~\ref{se:Feynrules}, and reads
\begin{align}
\Gamma_{W^+W^-\hat{\rho}}^{\mu\nu} (q_1,q_2) = 
\frac{g^2\tilde{v}}{4} \int_0^1 ds_1 \int_0^{s_1} ds_2 \, \, &
\left\{
\left[ \frac{\partial}{\partial q_\mu} \frac{\partial}{\partial q_\nu}  
\tilde{F} \left(q^2\right) \right]_{q=p + q_1 s_1 + q_2 s_2} \right. \\  
& \left.
\! + 
\left[ \frac{\partial}{\partial q_\mu} \frac{\partial}{\partial q_\nu}  
\tilde{F} \left(q^2\right) \right]_{q=q_1 s_1 + q_2 s_2} 
\right\} \, ,
\nonumber
\end{align}
where $q_1$ and $q_2$ are the W-boson momenta and $p=-q_1-q_2$ is the 
incoming momentum of the Higgs boson. This Feynman rule looks simple 
but the integrals cannot be generally performed without further knowledge 
on $\tilde{F}(p^2)$. However, for a two-to-two scattering process the 
longitudinal polarization vectors of the gauge bosons can be 
expressed in a rather simple way in terms of the gauge-boson momenta. 
For example, one finds
\begin{equation}
\label{eq:epsilon:expansion}
\epsilon_{L}^\mu (q_1)= \frac{1}{m_W \beta } 
\left( q_1^\mu - \frac{2 m_W^2}{s} (q_1+q_2)^\mu \right) 
= \frac{1}{m_W} \left( q_1^\mu - \frac{2m_W^2}{s} q_2^\mu \right) +
\mathcal{O}\!\left(\frac{q_{1,2}^\mu}{m_W} \frac{m_W^4}{s^2}\right),
\end{equation}
where $\beta = \sqrt{1 - 4 m_W^2/s}$ and $\vec{q}_1 = -\vec{q}_2$ 
has been used to establish the relation. Since~in two-to-two scattering 
the three-momenta of the final-state particles satisfy 
$\vec{q}_3= -\vec{q}_4$, an equivalent relation is valid for all four 
polarization vectors. Hence, for the scattering of longitudinal gauge bosons, 
it is sufficient to compute 
$q_{i\mu} q_{j\nu} \Gamma_{W^+W^-\hat{\rho}}^{\mu\nu}$. Moreover, 
at high energies, the leading approximation to the polarizations vector 
is proportional to the momentum of the corresponding gauge boson, while 
the next-to-leading approximation involves the other initial- or 
final-state momentum. 

Focusing on the leading approximation to the polarization vectors and the 
resulting leading terms in the amplitude, the vertex function contracted 
with the momenta of the gauge bosons can be expressed in terms of
derivatives with respect to the integration variables, which parameterize 
the Wilson line, 
\begin{align}
q_{1\mu} q_{2\mu} \Gamma_{W^+W^-\hat{\rho}}^{\mu\nu} (q_1,q_2)  = 
\frac{g^2\tilde{v}}{4} \int_0^1 ds_1 \int_0^{s_1} ds_2 \, \, 
\frac{\partial}{\partial s_1} \frac{\partial}{\partial s_2}  &
\left[ \tilde{F} \left((q_1 s_1 + q_2 s_2)^2\right) \right. \\ & + \left. \tilde{F} \left((q_1 (s_1-1) + q_2 (s_2-1))^2\right) \right]
\, .
\nonumber
\end{align}
After combining the two terms by a straight-forward variable 
transformation, the integrals can be done and yield
\begin{align}
q_{1\mu} q_{2\mu} \Gamma_{W^+W^-\hat{\rho}}^{\mu\nu} (q_1,q_2)  = &
\frac{g^2\tilde{v}}{4}\left[\tilde{F}\left((q_1 + q_2)^2\right) 
+ \tilde{F}(0) - 2 \tilde{F}(m_W^2) \right]\, .
\end{align}
The corresponding leading piece of the amplitude reads
\begin{equation}
\mathcal{M}_\mathrm{UH}^\mathrm{L,tot} =
\mathcal{M}_\mathrm{UH}^\mathrm{L}(s)+\mathcal{M}_\mathrm{UH}^\mathrm{L}(t)\, ,
\end{equation}
where
\begin{align}
\label{eq:WW-SKanal-Unhiggs-WL}
\mathcal{M}_\mathrm{UH}^\mathrm{L}(s) =& 
-i \frac{g^4\tilde{v}^2}{16m_W^4} \frac{1}{\tilde{F}(s)-\tilde{F}(M_H^2)}
\left[\tilde{F}(s) - \tilde{F}(0)\right]^2 \, 
\end{align}
and we have used $\tilde{F}(m_W^2)=\tilde{F}(0)$ since the difference is 
$\mathcal{O}(g^2)$ in terms which do not grow with energy.
We are interested in models in which $\tilde{F}(s)$ grows with energy but 
not faster than $s$, cf. \refequation{eq:unhigss_kernel}. 
Hence, the potentially unitarity-violating terms from the diagrams in 
Figure~\ref{fig:SM_Higgs_graphs} are given by
\begin{align}
\mathcal{M}_\mathrm{UH}^\mathrm{L,tot} =& -i \frac{g^4\tilde{v}^2}{16m_W^4} 
\left[\tilde{F}(s) + \tilde{F}(t)\right] + \mathcal{O}(s^0) \, .
\end{align}
Only in the SM limit, in which \ $\tilde{F}(s) + 
\tilde{F}(t) \to s+t +\mathcal{O}(s^0)$, the terms rising with energy
cancel those from the gauge sector. 

\begin{figure}
\begin{center}
\includegraphics[scale=.5]{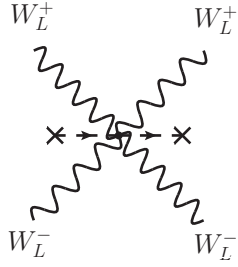}
\end{center}
\caption{\label{fig:4W_Unhiggs} Four-gauge-boson two-Unhiggs vertex 
diagram. The crosses indicate that a
Higgs boson has been replaced by its vev. }
\end{figure}

As has already been noted for the Mandelstam 
prescription~\cite{Stancato:2008mp}, the diagram shown in 
Figure~\ref{fig:4W_Unhiggs} contributes, which has no SM counterpart. 
Concentrating again on the leading terms of the polarization vectors, 
the corresponding vertex function contracted with momenta reads
\begin{align}
q_{1\mu} q_{2\nu} q_{3\rho} q_{4\sigma} 
\Gamma_{4G}^{\mu\nu\rho\sigma} (q_1,q_2,q_3,q_4)
= \,\,& 
\frac{g^4\tilde{v}^2}{8}  \int_0^1 ds_1 \int_0^{s_1} ds_2 \int_0^{s_2} ds_3 
\int_0^{s_3} ds_4 \\ 
& \hspace*{-5cm}
\Bigg[ \frac{\partial^{4}}{\partial s_1 \partial s_2 \partial s_3 
\partial s_4}
\tilde{F}\left((q_1s_1+q_2s_2+q_3s_3+q_4s_4)^2\right)
+ (q_1\leftrightarrow q_3) +  (q_2\leftrightarrow q_4) + 
\left(\begin{array}{c}q_1\leftrightarrow q_3 
\\q_2\leftrightarrow q_4\end{array}\right)\Bigg]\, .
\nonumber
\end{align}
Reordering integrations according to 
\begin{equation}
\int_0^1 ds_1 \int_0^{s_1} ds_2 \int_0^{s_2} ds_3 \int_0^{s_3} ds_4 = \int_0^1 ds_1 \int_0^{s_1} ds_3 
\int_{s_3}^{s_1} ds_2 \int_0^{s_3} ds_4 \, ,
\end{equation}
the last two integrals can be performed. In the remaining 
two-fold integral all terms in the integrand are either total derivatives 
with respect to one of the integrations or they depend only on a 
single integration variable after performing a suitable variable 
transformation. Hence, the result can be given in terms of 
one-dimensional integrals and the corresponding leading amplitude reads
\begin{equation}
\mathcal{M}_\mathrm{4G}^\mathrm{L,tot} = 
i \, \frac{q_{1\mu}}{m_W} \frac{q_{2\nu}}{m_W} \frac{q_{3\rho}}{m_W} 
\frac{q_{4\sigma}}{m_W} 
\, \Gamma_{4G}^{\mu\nu\rho\sigma}(q_1,q_2,q_3,q_4)  
= \mathcal{M}_\mathrm{4G}^\mathrm{L}(s,t) + 
\mathcal{M}_\mathrm{4G}^\mathrm{L}(t,s) \, ,
\label{eq:WW-4W-Unhiggs-WL}
\end{equation}
where
\begin{align}
\mathcal{M}_\mathrm{4G}^\mathrm{L}(s,t)					
= i \frac{g^4\tilde{v}^2}{8m_W^4} & \Bigg[4 s \, 
\frac{\tilde{F}(m_W^2)-\tilde{F}(0)}{m_W^2} 
+ \frac{1}{2} \left[\tilde{F}(s) -\tilde{F}(0)\right] 
\\&\hspace*{-2.5cm}
+ \left(\frac{s}{2} + t\right)\left[\frac{\tilde{F}(s)-\tilde{F}(0)}{s} 
- 2\int_0^1 \! dx \, \tilde{F}^{(1)}(s x^2)\right]
- \left(t + 2 s\right) \int_0^1 \! dx \,
 \tilde{F}^{(1)}\left(t x(x-1)\right)\Bigg]\,,
\nonumber 
\end{align}
and we have neglected $m_W^2$ compared to $s$ or $t$ in factors  
multiplying $\tilde{F}(p^2)$ as well as in its arguments. In contrast to 
the Mandelstam prescription, the contributions from this
diagram do not cancel all the unitarity-violating terms rising with 
energy in the other diagrams. The last term even contains a new 
integrand which is not present in any of the previously calculated 
amplitudes.

\begin{figure}
\begin{center}
\includegraphics[scale=.5]{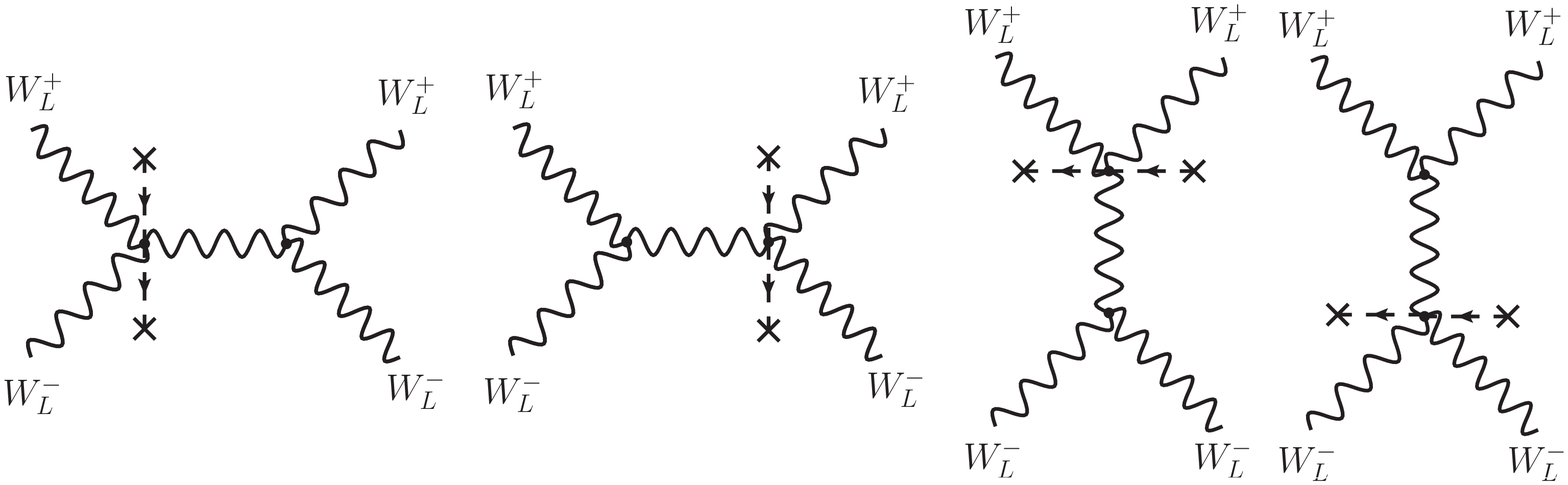}
\end{center}
\caption{\label{fig:Mixed_SM_Unhiggs_graphs} Additional graphs involving  
the two-Unhiggs triple gauge-boson 
vertex in combination with the SM triple gauge-boson vertex. }	
\end{figure}

There is yet another class of diagrams, which can be shown not to 
contribute in the Mandelstam case, but turns out to be essential for the 
unitarity cancellation in the Wilson-line model. The relevant vertex
results from the coupling of two Unhiggs fields with three gauge bosons 
when expanding the Wilson line to third order. The corresponding 
diagrams are shown in Figure~\ref{fig:Mixed_SM_Unhiggs_graphs}. Since we 
are interested only in the terms which grow with
energy, the masses in the propagators can be neglected such that we can 
calculate directly the exchange of the neutral SU(2) boson instead of a 
Z-boson and a photon.

The leading amplitude for the left diagram in 
Figure~\ref{fig:Mixed_SM_Unhiggs_graphs}, again replacing the four 
polarization vectors by the corresponding momenta, is given by
\begin{align}
\mathcal{M}^\mathrm{L}_{\mathrm{3G},1} =
& -i \frac{g^4\tilde{v}^2}{16m_W^4} \left(q_3-q_4\right)_\rho 
q_{1\mu} q_{2\nu}\\&
\times \int_0^1 ds_1 \int_0^{s_1} ds_2 \int_0^{s_2} ds_3 \,
\bigg[ \tilde{F}^{\mu\nu\rho} (s_1,s_2,s_3)
-  \tilde{F}^{\mu\nu\rho} (s_1,s_3,s_2)
+  \tilde{F}^{\mu\nu\rho} (s_2,s_3,s_1)\bigg] \, ,
\nonumber 
\end{align}
where
\begin{align}
\tilde{F}^{\mu\nu\rho}(s_1,s_2,s_3) =& 
\left.\frac{\partial^{3}}{\partial q_{\mu} \partial q_{\nu} \partial q_{\rho}}
\tilde{F}\left(q^2\right)\right|_{q=q_1s_1+q_2s_2+(q_3+q_4)s_3} \, .
\end{align}
The denominator of the propagator has canceled against a term resulting 
from the algebra of the SM three-gauge-boson vertex and the gauge-dependent 
terms of the propagator do not contribute. The three terms in the 
Unhiggs vertex result from the path ordering of the fields
and can be derived from the general result in Appendix~\ref{se:Feynrules} 
by setting the Higgs fields to their vev and making the SU(2) gauge 
fields explicit. Since $\tilde{F}^{\mu\nu\rho}(p^2)$ is again contracted 
with momenta, we can rewrite all 
derivatives as derivatives with respect to the integration variables, e.g.\ 
\begin{equation}
q_{1\mu} q_{2\nu} q_{3\rho} \tilde{F}^{\mu\nu\rho}(s_1,s_2,s_3) = 
\frac{\partial}{\partial s_1} \frac{\partial}{\partial s_2} 
\left[\frac{\partial}{\partial s_3} 
\tilde{F}\left((q_1s_1+q_2s_2+q_3s_3+q_4s_4)^2\right)
\right]_{s_4=s_3} \, .
\end{equation}
The six three-fold integrals for each of the four diagrams 
(most of them related by permutation of the momenta) can be again reduced 
to one-fold integrals and the combined 
result reads
\begin{equation}
\mathcal{M}^\mathrm{L,tot}_\mathrm{3G} = 
\mathcal{M}^\mathrm{L}_{\mathrm{3G}}(s,t)+ 
\mathcal{M}^\mathrm{L}_{\mathrm{3G}}(t,s)\, ,
\end{equation}
where
\begin{align}
\label{eq:WW-SKanal-Mischgraph}
\mathcal{M}^\mathrm{L}_{\mathrm{3G}}(s,t) =
-i \frac{g^4\tilde{v}^2}{8m_W^4}& 
\Bigg\{2 \left(s+ 2 t\right) \frac{\tilde{F}(m_W^2)-\tilde{F}(0)}{m_W^2}
\\& \hspace*{-3cm}
 - \left(s + 2t\right)\int_0^1 dx \tilde{F}^{(1)} 
\left(s(x-1)x\right)
+ \left(s + 2t\right)\Bigg[ \frac{\tilde{F}\left(s\right) - \tilde{F}(0)}{s}
-2\int_0^1 dx \tilde{F}^{(1)}\!\left(s x^2\right)\Bigg]\Bigg\}\,. 
\nonumber
\end{align}
As before, we neglected $m_W^2$ compared to terms of order $s$.

\begin{figure}
\begin{center}
\includegraphics[scale=.5]{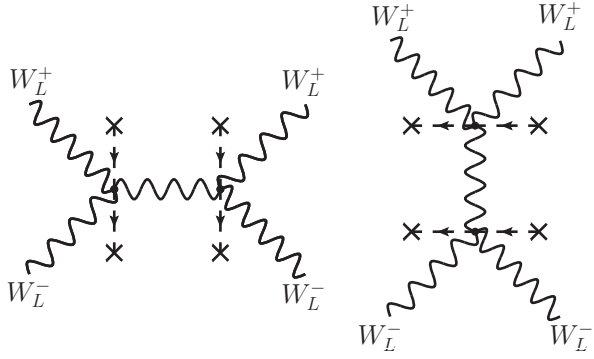}
\end{center}
\caption{\label{fig:Unhiggs_Unhiggs_graphs} Additional graphs resulting 
from two two-Unhiggs triple gauge-boson vertices.}	
\end{figure}

The last class of diagrams to consider is displayed in 
Figure~\ref{fig:Unhiggs_Unhiggs_graphs}. In order to show that these 
two diagrams do not contribute to the unitarization of W-boson scattering,
we have to further investigate the vertices. Each vertex is connected 
to two external lines and the corresponding Lorentz indices are again 
contracted with the momenta of the external bosons in the leading 
approximation. With one open Lorentz index, we now have to evaluate
\begin{eqnarray}
 q_{1\mu} q_{2\nu} \Gamma_{3G}^{\mu\nu\rho} &\propto & 
\, q_{1\mu} q_{2\nu}
\int_0^1 ds_1 \int_0^{s_1} ds_2 \int_0^{s_2} ds_3 \,
\nonumber\\
&&
\bigg[ \tilde{F}^{\mu\nu\rho} (s_1,s_2,s_3)
- \tilde{F}^{\mu\nu\rho} (s_1,s_3,s_2)
+ \tilde{F}^{\mu\nu\rho} (s_2,s_3,s_1)\bigg] \, .
\end{eqnarray}
The second term is easily calculated because the integrand is a total 
derivative with respect to both  the $s_1$ and $s_3$ integrations. Hence, 
the integrand of the remaining $s_2$ integration is proportional to 
$\tilde{F}^{(1)}(p^2)$ and a sum of uncontracted momentum vectors. Since we 
are only concerned with kernels for which $\tilde{F}^{(1)}(p^2)$ goes to 
zero when $p^2$ becomes large, these terms  rise with energy at most 
like a single momentum. For the first and the third term, the integrand  
is a total derivative with respect to the $s_2$ integration. For example,
the first term results in
\begin{align}
\label{eq:suppression}
q_{1\mu} q_{2\nu} \int_0^1 ds_1 \int_0^{s_1} ds_3 & 
\int_{s_3}^{s_1} ds_2 \,\tilde{F}^{\mu\nu\rho} (s_1,s_2,s_3) = \\ 
\nonumber & \hspace*{-2cm}
q_{1\mu} \int_0^1 ds_1 \int_0^{s_1} ds_3 \,
\tilde{F}^{(2)}\left((s_1-s_3)^2(q_1+q_2)^2\right) 
4 (s_1-s_3)^2(q_1+q_2)^{\mu} (q_1+q_2)^{\rho} \\  
\nonumber & \hspace*{-2cm} 
- q_{1\mu} \int_0^1 ds_1 \!\!\int_0^{s_1} ds_3 \,
\tilde{F}^{(2)}\left((s_1-s_3)^2q_1^2\right) 
4 (s_1-s_3)^2 q_1^{\mu} q_1^{\rho} 
+ \ldots \, ,
\end{align}
where we have used momentum conservation in the vertex and omitted terms 
which only contain first derivatives of $\tilde{F}$ and are not dangerous 
at high energies. Changing integration variables to $(s_1-s_3)$ and 
integrating by parts, we again derive a result which contains 
only first derivatives of $\tilde{F}$. The first term again rises with 
energy only like a single momentum, because partial integration yields a 
large suppression factor $(q_1+q_2)^2\sim s$ in the denominator. The second 
term is potentially more dangerous at high energies because
one finds a factor $q_1^2=m^2_W$ in the denominator and three momenta in 
the numerator. This behaviour follows from \refequation{eq:suppression} 
since the argument of $\tilde{F}^{(2)}\left((s_1-s_3)^2 m_W^2\right)$ does 
not rise with energy and cannot suppress the integrand at large energies 
when the momenta in the numerator become large. However, as also seen 
from the equation, dotting $q_1$ into the vertex converts momentum 
factors into the small quantity $m_W^2$. It is a generic feature 
that the terms in the vertices which potentially rise fastest with 
energy are tamed by explicit factors of $m_W^2$ in the numerator. 
Hence, the vertex contracted with the external momenta only rises 
with energy like a single momentum. It follows that the diagrams in 
Figure~\ref{fig:Unhiggs_Unhiggs_graphs}, which are built by two of these
contracted vertices and a propagator do not lead to  
unitarity-violating terms in the amplitude. The diagrams in 
Figure~\ref{fig:Mixed_SM_Unhiggs_graphs} do contribute because the SM
vertex contracted with external momenta rises like the third power of 
the centre-of-mass energy. 

Adding up all terms the complete leading amplitude from the SM 
and Unhiggs diagrams is given by
\begin{align}
\label{eq:WW-WilsonLinie-Ergebnis}
\mathcal{M}^{\mathrm{L,tot}} =& 
\,\,\mathcal{M}_{\mathrm{GS}} +
\mathcal{M}^{\mathrm{L,tot}}_{\mathrm{UH}} +
\mathcal{M}^{\mathrm{L,tot}}_{\mathrm{4G}} +
\mathcal{M}^{\mathrm{L,tot}}_{\mathrm{3G}} 
\\ \nonumber
=&\,\,i \frac{g^4\tilde{v}^2}{8m_W^4} 
\left[4 (s+t)\left(\int_0^1 dx \tilde{F}^{(1)}(m_W^2x^2)
- \frac{\tilde{F}(m_W^2)-\tilde{F}(0)}{m_W^2}\right)\right.\\&
+ \tilde{F}(0)-\tilde{F}(M_H^2)\nonumber \left.
- \frac{1}{2} 
\left(\frac{(\tilde{F}(0)-\tilde{F}(M_H^2))^2}{\tilde{F}(s)-\tilde{F}(M_H^2)} 
+ \frac{(\tilde{F}(0)-\tilde{F}(M_H^2))^2}{\tilde{F}(t)-\tilde{F}(M_H^2)}
\right)\right]\nonumber
+ \mathcal{O}\!\left( g^2 \right)\, ,
\end{align}
where we have kept all terms which rise with energy or do not go to zero 
in the limit of vanishing gauge couplings. Most of the potentially 
unitarity-violating terms in the different diagrams have indeed canceled. 
However, the terms in the first line still violate unitarity at high
energies. 

As we show in the following, the unitarity restoration occurs in the 
present non-local Higgs theory only when sub-leading 
terms in the longitudinal polarization 
vectors~\refequation{eq:epsilon:expansion} are included.
The critical diagrams are those from Figures~\ref{fig:4W_Unhiggs} 
and~\ref{fig:Mixed_SM_Unhiggs_graphs} from the 
triple and quartic gauge-boson vertex in the Unhiggs sector. 
Substituting the sub-leading term from~\refequation{eq:epsilon:expansion} in 
one of the four polarization vectors in the quartic gauge-boson vertex
diagram and summing over the four 
possibilities, the next-to-leading amplitude is given by
\begin{equation}
\hspace*{-0.3cm}
\mathcal{M}_\mathrm{4G}^\mathrm{NL,tot} = \frac{-2i}{s m_W^2} \left[
\left(q_{1\mu} q_{1\nu} + q_{2\mu} q_{2\nu} \right) q_{3\rho} q_{4\sigma} +
q_{1\mu} q_{2\nu} \left( q_{3\rho} q_{3\sigma}+
q_{4\rho} q_{4\sigma} \right)\right]
\Gamma_{4G}^{\mu\nu\rho\sigma}(q_1,q_2,q_3,q_4) \, .
\end{equation}
As for the leading amplitude two integrations inside the vertex can 
always be trivially performed. The discussion 
after \refequation{eq:suppression} shows that only terms including
$\tilde{F}^{(2)}((s_i-s_j)m_W^2)$ can contribute because of the extra 
factor $1/s$ in the next-to-leading amplitude. All these terms can be 
further integrated and yield
\begin{align}
\mathcal{M}_\mathrm{4G}^\mathrm{NL,tot}
= & \,\,i \frac{g^4\tilde{v}^2}{m_W^4} (s+t)\left(\int_0^1 dx 
\,\tilde{F}^{(1)}(m_W^2x^2)
- \frac{\tilde{F}(m_W^2)-\tilde{F}(0)}{m_W^2}\right) \, .
\end{align}
The equivalent analysis of the next-to-leading terms for the diagrams in 
Figure~\ref{fig:Mixed_SM_Unhiggs_graphs} results in 
\begin{equation}
\mathcal{M}_\mathrm{3G}^\mathrm{NL,tot} = 
\mathcal{M}_\mathrm{3G}^\mathrm{NL}(s,t)+
\mathcal{M}_\mathrm{3G}^\mathrm{NL}(t,s) \,
\end{equation}
where
\begin{equation}
\mathcal{M}_\mathrm{3G}^\mathrm{NL}(s,t) = 
-i \frac{g^4\tilde{v}^2}{2 m_W^4} (s+2t)\left(\int_0^1 dx 
\,\tilde{F}^{(1)}(m_W^2x^2)
- \frac{\tilde{F}(m_W^2)-\tilde{F}(0)}{m_W^2}\right) \, .
\end{equation}
The diagrams in Figure~\ref{fig:Unhiggs_Unhiggs_graphs} once again do 
not contribute to the terms rising with energy. There are a few terms 
appearing in the evaluation of $\mathcal{M}_\mathrm{4G}^\mathrm{NL,tot}$ 
and $\mathcal{M}_\mathrm{3G}^\mathrm{NL,tot}$ which cannot be rewritten in 
terms of one-fold integrals. Hence, it is not obvious for a general 
kernel that these terms do not contribute. However, when adding these 
terms from the different sets of diagrams we find that they cancel exactly. 

Finally, adding up all the leading and next-to-leading terms, all 
unitarity-violating terms cancel. That is, quite remarkably, the 
left-over terms 
in the second line of~\refequation{eq:WW-WilsonLinie-Ergebnis} are indeed 
canceled by those from the next-to-leading terms in the polarization 
vectors. Since next-to-next-to-leading terms in the expansion of the 
longitudinal polarization vectors are suppressed by
another factor of $m_W^2/s$ and cannot contribute to potentially 
unitarity-violating terms, this completes the explicit demonstration 
that high-energy unitarity is not spoiled by the non-local Higgs sector. 
We further note that non-leading pieces in all the diagrams related to the 
Unhiggs sector do not contribute at $\mathcal{O}(g^0)$, in contrast to 
non-leading pieces in the SM like diagrams which are already included. 
Hence, the final result for the amplitude reads
\begin{align}
\label{eq:final_amplitude}
\mathcal{M} \! =&
\frac{-i}{\tilde{F}^{(1)}(0)^2} \! \left[\,2 V^{(2)}(\tilde{v}^2/2) 
+ \tilde{v}^2 \,(V^{(2)}(\tilde{v}^2/2))^2
\! \left(\frac{1}{\tilde{F}(s)-\tilde{F}(M_H^2)} +
\frac{1}{\tilde{F}(t)-\tilde{F}(M_H^2)}\right)\,\right] \!\! +\mathcal{O}(g^2)\, ,
\end{align}
where we have used~\refequation{eq:Higgsmass} and the relations
between the different mass parameters 
$\tilde{m}^2_W=g^2\tilde{v}^2/4=\hat{m}^2_W/\tilde{F}^{(1)}(0)=
m_W^2/\tilde{F}^{(1)}(0) \times [1+\mathcal{O}(g^2)]$. 

\subsection{Comparison to minimal coupling}
\label{sec:comparison}

As we have shown in the previous section, unitarization of longitudinal 
W-boson scattering does not only take place for a minimally coupled 
action but also for an action based on the straight Wilson line. While
the cancellation of potentially unitarity violating terms is based in 
both cases on the non-trivial interplay of different diagrams, the 
cancellation for the straight Wilson line is more involved, and 
additionally involves the graphs in Figure~\ref{fig:Mixed_SM_Unhiggs_graphs}. 

The terms (\ref{eq:final_amplitude}) remaining in the limit of 
vanishing gauge couplings are identical for the minimally coupled and 
Wilson-line actions. Using the non-local kernel (\ref{eq:unhigss_kernel}) 
and the Higgs potential~(\ref{eq:Higgspotential}), our 
result~\refequation{eq:final_amplitude} matches the result in Eq.~(3.29)
of Ref.~\cite{Stancato:2008mp}. This coincidence of results is 
certainly not accidental and suggests that the Goldstone-boson equivalence 
theorem applies to non-local Higgs actions as we verify below. The 
left-over terms in~\refequation{eq:final_amplitude} imply as usual constraints 
on the Higgs potential, since the energy-independent terms must not 
be too large to satisfy unitarity. However, since there is no difference 
to the minimal-coupling result discussed in Ref.~\cite{Stancato:2008mp}, 
we do not repeat the corresponding analysis of constraints on $d$ 
and $\mu$ here.

\subsection{Goldstone-boson equivalence theorem}
\label{sec:equivalence_theorem}

In the SM the Goldstone-boson equivalence 
theorem~\cite{Cornwall:1974km,Lee:1977eg,Chanowitz:1985hj} for 
longitudinal W-boson scattering reads
\begin{align}
\label{eq:equivalence_theorem}
&\mathcal{M} \left[W^{+,L} (p_1)+W^{-,L} (p_2) \to 
W^{+,L} (p'_1)+W^{-,L} (p'_2) \right]\\
& \quad = \, \mathcal{M} \left[\pi^{+}(p_1)+ \pi^{-}(p_2)  
\to \pi^{+}(p'_1)+ \pi^{-}(p'_2)\right]
+ \mathcal{O} \left(\frac{m_W}{E}\right)\,, 
\nonumber 
\end{align}
i.e.\ at high energies the longitudinal W boson scattering amplitude 
equals the scattering amplitude of the corresponding Goldstone bosons 
up to terms that vanish in the high-energy limit. In the following, we 
again consider the limit of vanishing gauge couplings, in which the terms 
of order $m_W/E$ vanish and the two amplitudes are predicted to 
agree precisely. 

In this work, we do not attempt a formal proof of the Goldstone-boson 
equivalence theorem for non-local Higgs-gauge theories but verify it 
by a tree-level calculation of the Goldstone scattering amplitude on the 
right hand side of \refequation{eq:equivalence_theorem}. That the 
theorem is likely to be valid in the present class of non-local theories 
can be understood by the following considerations: Since we are not 
considering corrections due to the gauge couplings, we can set 
$m_\mathrm{V}=\hat{m}_\mathrm{V}$. The gauge-fixing 
condition \refeq{eq:gauge_fixing_condition} is then  identical to SM 
$R_\xi$-gauge. Hence, the resulting Ward identity, which is the cornerstone 
of formal proofs~\cite{Chanowitz:1985hj,Gounaris:1986cr} of the equivalence 
theorem, can be expected to hold unmodified for the non-local Higgs model. 
Moreover, as shown above 
the propagators for the longitudinal gauge bosons and the 
Goldstone bosons only  differ by a factor of $\xi$ (as in the SM), so  
the Ward identity for connected Green functions can be translated into 
an identity for matrix elements, which yields the desired 
result.\footnote{Note that the normalization of the Goldstone fields 
chosen in~\refequation{eq:Goldstone_normalization} is essential in this
respect. Otherwise, non trivial $Z$-factors for the external Goldstone 
bosons would have to compensate the corresponding
factors in the Feynman rules for the explicit amplitude calculation 
below.}

The calculation of the amplitude for charged Goldstone-boson scattering 
$\pi^+\pi^-\to\pi^+\pi^-$ in the limit of vanishing gauge couplings 
depends only on the Higgs potential~\refeq{eq:higgspotential}. By expanding 
the Higgs potential to second order around the vev, the interaction terms 
relevant to charged Goldstone scattering at tree-level are given by
\begin{equation}
S \supset -\int d^4 x \,\frac{1}{2} \,
\frac{V^{(2)}(\tilde{v}^2/2)}{\tilde{F}^{(1)}(0)^2} 
\left( 2 v \,\rho \, \pi^+ \pi^- + (\pi^+ \pi^-)^2 \right)\, .
\end{equation}
The Feynman rules for the triple and quartic scalar interaction 
vertices can be deduced from this expression. The calculation of the 
tree-level Goldstone scattering amplitude is straight-forward 
resulting in 
\begin{align}
\label{eq:Goldstone_amplitude}
\mathcal{M} = & 
-i \left[\,2 \, \frac{V^{(2)}(\tilde{v}^2/2)}{\tilde{F}^{(1)}(0)^{2}} 
+ v^2 \left(\frac{V^{(2)}(\tilde{v}^2/2)}{\tilde{F}^{(1)}(0)^{2}}
\right)^{\!2}
\left(\frac{\tilde{F}^{(1)}(0)}{\tilde{F}(s)-\tilde{F}(M_H^2)} + 
\frac{\tilde{F}^{(1)}(0)}{\tilde{F}(t)-\tilde{F}(M_H^2)}\right)\right] \, .
\end{align}
The first term is due to the vertex with four charged Goldstone bosons. 
The remaining two stem from the $s$- and $t$-channel exchange of the  
physical Higgs boson, respectively. Using 
$v^2=\tilde{v}^2 \tilde{F}^{(1)}(0)$, the longitudinal WW scattering  
amplitude \refeq{eq:final_amplitude} and 
\refequation{eq:Goldstone_amplitude} indeed agree as they should, if the 
Goldstone-boson equivalence theorem holds.

\section{Summary}
\label{sec:summary}

The present work has been motivated by the observation made in 
Ref.~\cite{Stancato:2008mp} that the standard mechanism of electroweak 
symmetry breaking by the Higgs field can be made to work  
when the Higgs action is non-local. While the specific form 
of the action \refeq{eq:snonlocal1}, and the assumption that only the 
Higgs sector is non-local may be hard to justify from the 
viewpoint of model-building, the fact that a radical modification  
from the standard set-up such as relinquishing locality appears to 
be consistently unitarizing the theory at energies above the 
symmetry-breaking scale is certainly interesting. 

In previous work a minimal-coupling prescription has been employed to 
render the theory gauge-invariant, but this is far from 
unique when the action is non-local and cannot be expanded into a 
series of local operators of increasing dimension. Here we considered 
non-local theories with general kernels and gauge invariance 
maintained by a straight Wilson line extended between the positions 
of the Higgs fields. One of our main findings is that longitudinal 
WW scattering unitarizes, but as a consequence of diagrammatic 
cancellations that are remarkably non-trivial compared to the 
SM and even to the minimally-coupled non-local Higgs model. We 
verified that the Goldstone-boson equivalence theorem is 
fulfilled and suggest that its formal proof should be extensible 
to the non-local case.

For particle physics model building the model does not look promising, 
however. Contrary to Ref.~\cite{Stancato:2008mp}, we find that the 
quantum corrections to the Higgs mass are not reduced (unless the 
dimension of the Higgs field is very close to the pathological 
limit $d=2$), when expressed in terms of known low-energy 
parameters, and hence the model does not reduce the (little) hierarchy 
problem in the sense of allowing a larger cut-off than the SM for 
the same amount of cancellations in the Higgs self-energy. Furthermore, 
unlike the minimally-coupled theory, 
in general there exist tree-level corrections to the transverse 
gauge-boson propagators, leading to unacceptably large values of 
the $S$-parameter and the W-mass shift, unless the scale $\mu$ of 
non-locality is significantly larger than the electroweak scale, 
or $d$ is close to the SM limit $d=1$. 
But sufficiently below the scale $\mu$ the non-local Higgs sector 
is indistinguishable from the standard local implementation. 

\subsubsection*{Acknowledgement}
We thank M.~Luty, J.~Rohrwild and J.~Terning for useful discussions. 
MB thanks the Kavli Institute for Theoretical Physics 
at UC Santa Barbara for hospitality, while part of this work 
was done. 
This work is supported in part by the Gottfried Wilhelm 
Leibniz programme of the Deutsche Forschungsgemeinschaft (DFG) and
by the National Science Foundation under Grant No. NSF PHY05-51164.

\begin{appendix}
\section{Generic interactions}
\label{se:Feynrules}

In this appendix the interaction terms for the straight Wilson-line 
coupling of the non-local Higgs sector to the gauge fields are derived. 
Expanding the path-ordered exponential, 
the action~\refeq{eq:wirkungortsraum} is given by 
\begin{align}
S \supset & \int d^4x \int d^4y \,F(x-y) 
\\
&\times \sum_{n=1}^\infty \,(-ig)^n \int_y^x dz_{1,\nu_1}
\int_y^{z_1} dz_{2,\nu_2} \cdots \int_y^{z_{n-1}} dz_{n,\nu_n} 
\,\phi^\dagger (x) \, A^{\nu_1} (z_1)\cdots A^{\nu_n} (z_n) \, \phi(y) \, ,
\nonumber
\end{align}
where $A^{\nu_n} (z_n)= \tau^{a_n} A^{\nu_n}_{a_n} (z_n)$ is the  
matrix-valued gauge field of the gauge group under consideration.
Note that the definition of the path-ordered exponential does not include 
factors of $1/n!$. For the straight
Wilson line, the path is parameterized by 
$z^{\nu}_n=(x-y)^{\nu} s_n+y^{\nu}$ yielding
\begin{align}
S \supset & \int d^4x \int d^4y \,F(x-y) \,\sum_{n=1}^\infty \,(-ig)^n 
\int_0^1  ds_{1}\,(x-y)_{\nu_1} \\
&\times
\int_0^{s_1}  ds_{2}\,(x-y)_{\nu_2} \cdots 
\int_0^{s_{n-1}}  ds_{n}\,(x-y)_{\nu_n} \,
\phi^\dagger (x) \, A^{\nu_1} (s_1)\cdots A^{\nu_n} (s_n) \, \phi(y) \, .
\nonumber
\end{align}
In momentum space the action is then given by 
\begin{align}\nonumber
S \supset & \sum_{n=1}^\infty \,(-ig)^n 
\int\frac{d^4q}{(2\pi)^4}\frac{d^4p_1}{(2\pi)^4}\frac{d^4p_2}{(2\pi)^4}
\frac{d^4q_1}{(2\pi)^4}\cdots \frac{d^4q_n}{(2\pi)^4} \,
\tilde{F}(q^2) \,\phi^\dagger (p_1) A^{\nu_1} (q_1) \cdots A^{\nu_n} (q_n) \, 
\phi(p_2)\\
&\times  
\frac{i\partial}{\partial q^{\nu_1}} \cdots 
\frac{i\partial}{\partial q^{\nu_n}} 
\int \! d^4x \! \int \! d^4y \! \int_0^1 \!\!\! ds_{1} \cdots \! \int_0^{s_{n-1}} \!\!\!\!\!ds_{n} \,
e^{-ix(q-p_1+\sum_{k=1}^n s_k q_k)} e^{+iy(q-p_2+\sum_{k=1}^n (s_k-1) q_k)}\, .
\end{align}
Performing the integrations over the variables $x$ and $y$, using partial 
integration and finally performing the $q$-integration, we find 
\begin{align}
S \supset & \sum_{n=1}^\infty \,g^n 
\int\frac{d^4p_1}{(2\pi)^4}\frac{d^4p_2}{(2\pi)^4}
\frac{d^4q_1}{(2\pi)^4}\cdots\frac{d^4q_n}{(2\pi)^4} \,
\phi^\dagger (-p_1) A^{\nu_1} (q_1) \cdots A^{\nu_n} (q_n) \, \phi(p_2) \\
&\times  
(2\pi)^4\delta^{(4)}\left(p_1+p_2+\sum_{k=1}^n q_k\right) 
\int_0^1ds_{1}\, \cdots \int_0^{s_{n-1}} ds_{n} \,\, 
\left. \frac{\partial}{\partial q^{\nu_1}} 
\cdots \frac{\partial}{\partial q^{\nu_n}} \tilde{F}(q^2) 
\right|_{q=p_1+\sum_{k=1}^n s_k q_k} \, .
\nonumber
\end{align}
This expression allows us to read off the Feynman rule for a vertex 
with two Higgs and any number of gauge fields. For the specific case 
of the electroweak gauge group, where one is interested in the 
interactions of the physical Higgs field and the Goldstone modes with
photons, W- and Z-bosons, one can easily make the group factors, 
hidden in the matrix notation for the gauge fields and the Higgs doublets,
explicit. Feynman rules involving the Higgs vacuum expectation value 
are obtained by setting the momentum of the corresponding Higgs field to 
zero.

\end{appendix}



\begin{thebibliography}{99}

\bibitem{Georgi:2007ek}
  H.~Georgi,
  Phys.\ Rev.\ Lett.\  {\bf 98 } (2007)  221601,
  hep-ph/0703260.

\bibitem{Stancato:2008mp}
  D.~Stancato, J.~Terning,
  JHEP {\bf 0911} (2009)  101,
  arXiv:0807.3961 [hep-ph].

\bibitem{Falkowski:2008yr}
  A.~Falkowski, M.~Perez-Victoria,
  Phys.\ Rev.\  {\bf D79 } (2009)  035005,
  arXiv:0810.4940 [hep-ph].

\bibitem{Cacciapaglia:2008ns}
  G.~Cacciapaglia, G.~Marandella, J.~Terning,
  JHEP {\bf 0902 } (2009)  049,
  arXiv:0804.0424 [hep-ph].

\bibitem{Falkowski:2008fz}
  A.~Falkowski, M.~Perez-Victoria,
  JHEP {\bf 0812 } (2008)  107,
  arXiv:0806.1737 [hep-ph].

\bibitem{Randall:1999ee}
  L.~Randall, R.~Sundrum,
  Phys.\ Rev.\ Lett.\  {\bf 83 } (1999)  3370-3373,
  hep-ph/9905221.

\bibitem{Mandelstam:1962mi}
  S.~Mandelstam,
  Annals Phys.\  {\bf 19 } (1962)  1-24.

\bibitem{Galloway:2008jn}
  J.~Galloway, D.~Martin, D.~Stancato,
  arXiv:0802.0313 [hep-th].

\bibitem{Ilderton:2008ab}
  A.~Ilderton,
  Phys.\ Rev.\  {\bf D79 } (2009)  025014,
  arXiv:0810.3916 [hep-th].

\bibitem{LewisLicht:2008mq}
  A.~L.~Licht,
  arXiv:0802.4310 [hep-th].

\bibitem{Licht:2008km}
  A.~L.~Licht,
  arXiv:0806.3596 [hep-th].

\bibitem{LewisLicht:2008mv}
  A.~L.~Licht,
  arXiv:0805.3849 [hep-th].

\bibitem{Peskin:1991sw}
  M.~E.~Peskin, T.~Takeuchi,
  Phys.\ Rev.\  {\bf D46 } (1992)  381-409.

\bibitem{Nakamura:2010zzi}
  K.~Nakamura {\it et al.} [ Particle Data Group Collaboration ],
  J.\ Phys.\ G {\bf G37 } (2010)  075021.

\bibitem{Cornwall:1974km}
  J.~M.~Cornwall, D.~N.~Levin, G.~Tiktopoulos,
  Phys.\ Rev.\  {\bf D10 } (1974)  1145.

\bibitem{Lee:1977eg}
  B.~W.~Lee, C.~Quigg, H.~B.~Thacker,
  Phys.\ Rev.\  {\bf D16 } (1977)  1519.

\bibitem{Chanowitz:1985hj}
  M.~S.~Chanowitz, M.~K.~Gaillard,
  Nucl.\ Phys.\  {\bf B261 } (1985)  379.

\bibitem{Gounaris:1986cr}
  G.~J.~Gounaris, R.~K\"ogerler, H.~Neufeld,
  Phys.\ Rev.\  {\bf D34 } (1986)  3257.

\end{thebibliography}
\end{document}